\documentclass[aps,pra,10pt,twocolumn,showpacs]{revtex4-1}
\usepackage{amsmath}
\usepackage{graphicx}
\usepackage{dcolumn}
\begin{document}
\title{Intercombination Effects in Resonant Energy Transfer}
\author{C L Vaillant}
\email{c.l.j.j.vaillant@durham.ac.uk}
\author{R M Potvliege}
\author{M P A Jones}
\email{m.p.a.jones@durham.ac.uk}
\affiliation{Department of Physics, Joint Quantum Centre (JQC) Durham-Newcastle, Durham University, South Road, Durham DH1 3LE, United Kingdom}
\begin{abstract}
We investigate the effect of intercombination transitions in excitation hopping processes such as those found in F\"orster resonance energy transfer. Taking strontium Rydberg states as our model system, the breakdown of $LS$-coupling leads to weakly allowed transitions between Rydberg states of different spin quantum number. We show that the long-range interactions between two Rydberg atoms can be affected by these weakly allowed spin transitions, and the effect is greatest when there is a near-degeneracy between the initial state and a state with a different spin quantum number. We also consider a case of four atoms in a spin chain, and show that a spin impurity can resonantly hop along the chain. By engineering the many-body energy levels of the spin-chain, the breakdown of $LS$ coupling due to inter-electronic effects in individual atoms can be mapped onto a spatial separation of the total spin and the total orbital angular momentum along the spin chain.
\end{abstract}

\maketitle

\section{Introduction}
Non-radiative exchanges of energy mediated by dipole-dipole interactions play a crucial role in a variety of processes, ranging from photosynthesis in natural biological systems~\cite{Cheng2009} to highly efficient light emission in organic devices~\cite{Baldo2000}. The usual requirements for a significant energy transfer are an electric dipole-dipole interaction between donor and acceptor molecules, and a near degeneracy between the initial and final states that ensures the process is always resonant~\cite{Scholes2003}. Resonant energy transfer has also been extensively studied in atomic physics, where these conditions are easily met. For example, dipole-dipole interactions in dense, optically excited samples can lead to cooperative Lamb shifts~\cite{Keaveney2012,Rohlsberger2010}. By using Rydberg states, rather than low-lying electronic states, the strength of the interaction can be increased by many orders of magnitude, leading to energy exchange over macroscopic distances~\cite{Ravets2014,Barredo2015,Vogt2007,Ryabtsev2010,Vogt2006,vanDitzhuijzen2008,Gunter2013,Anderson1998,Mourachko1998}.

It is often assumed, in agreement with the electric dipole selection rules, that only states with the same value of the total electron spin quantum number $S$ are coupled through dipole-dipole interactions. This assumption is justified for interactions between alkali atoms, since $S=1/2$ for all the states relevant in this context. However, in atomic and molecular systems with more than one valence electron, $S$ is at best an approximately good quantum number owing to inter-electronic interactions and spin-orbit coupling. Intercombination transitions in resonant energy transfer have been considered in molecules~\cite{Forster1959} and quantum dots~\cite{Govorov2005}, and experimental observations~\cite{Ermolaev1963,Bennett1964} include important applications in efficient organic light-emitting devices~\cite{Baldo2000}. In general, strong coupling to a bath of rotational, vibrational and motional states prevents the observation of coherent transport in these systems.

In this article, we examine the impact of singlet-triplet mixing on the long-range interaction between isolated ultracold atoms in the excitonic regime \cite{Clegg2010}, where dephasing due to effects such as molecular motion is sufficiently small that the transport may be considered coherent. Our treatment may also find applications in molecular systems where coherent transfer plays a role. First we consider the effect of intercombination transitions on non-resonant van der Waals-type interactions, and second the effect of spin mixing on resonant hopping processes. We find that even in systems where spin mixing is relatively weak, near degeneracies between donor and acceptor states such as those that occur in F\"orster resonances can lead to novel transport effects which depend on the value of the spin quantum number $S$ --- here we refer to the actual electronic spin, rather than, e.g., to 2-level excitation mapped to a pseudo-spin or to the spin magnetic quantum number $M_S$ (i.e., the spin orientation). Excitation transport in atomic lattices is an area of great current interest due to potential applications of cold-atom model systems to condensed matter research \cite{Bloch2008,Lewenstein2007}. Here we consider a one-dimensional lattice of strontium atoms in $5snd$ Rydberg states, thus adding strong inter-site interactions and extending the study of many-body systems beyond the two-level Hubbard model in cold atoms~\cite{Lukin2001,Bouchoule2002,Lesanovsky2011}. Ultra-cold Rydberg gases of divalent atoms are of growing interest in atomic physics~\cite{Millen2010,Gil2014,Ye2013,McQuillen2013,Mukherjee2011,Lochead2013}, and as well as systems where the precise details of the electronic wave function are known~\cite{Vaillant2014a, Esherick1977a, Aymar1987, Hiller2014,Topcu2014,Millen2011}, they provide a route to precise control of the inter-particle spacing via optical lattices or tweezer arrays~\cite{Ovsiannikov2011,Ido2003,Nogrette2014,Piotrowicz2013,Schlosser2011}. 

\begin{figure}
\centering
\includegraphics{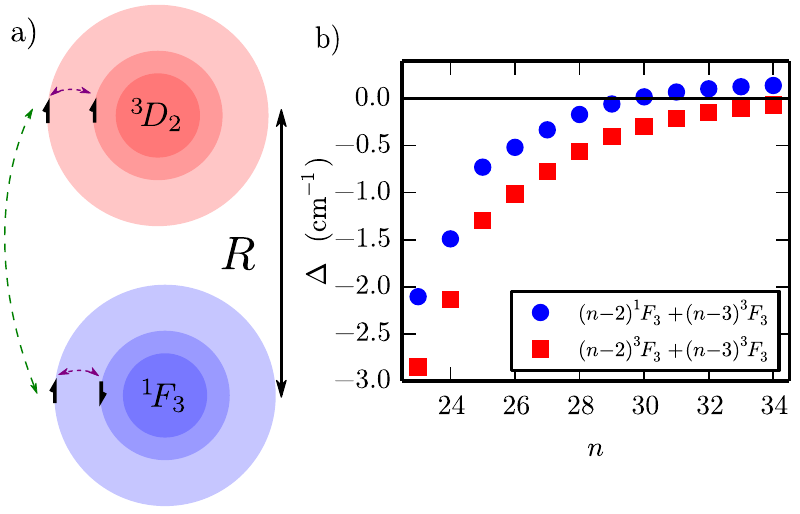}
\caption{\label{mainfigure} (color online) (a) Two Rydberg atoms separated by a distance $R$ are prepared in different spin states. Interactions between core and valence electrons (purple dashed arrow) leads to a breakdown of $LS$ coupling, enabling an otherwise forbidden dipole-dipole interaction to take place (green dashed arrow). (b) The energy separation between the $|30\, ^3\!D_2, 30\, ^3\!D_2\rangle$ pair states and the energetically closest spin allowed (red squares) and spin forbidden (blue circles) final pair states, showing a spin-forbidden resonance.}
\end{figure}

\begin{figure}
\centering
\includegraphics{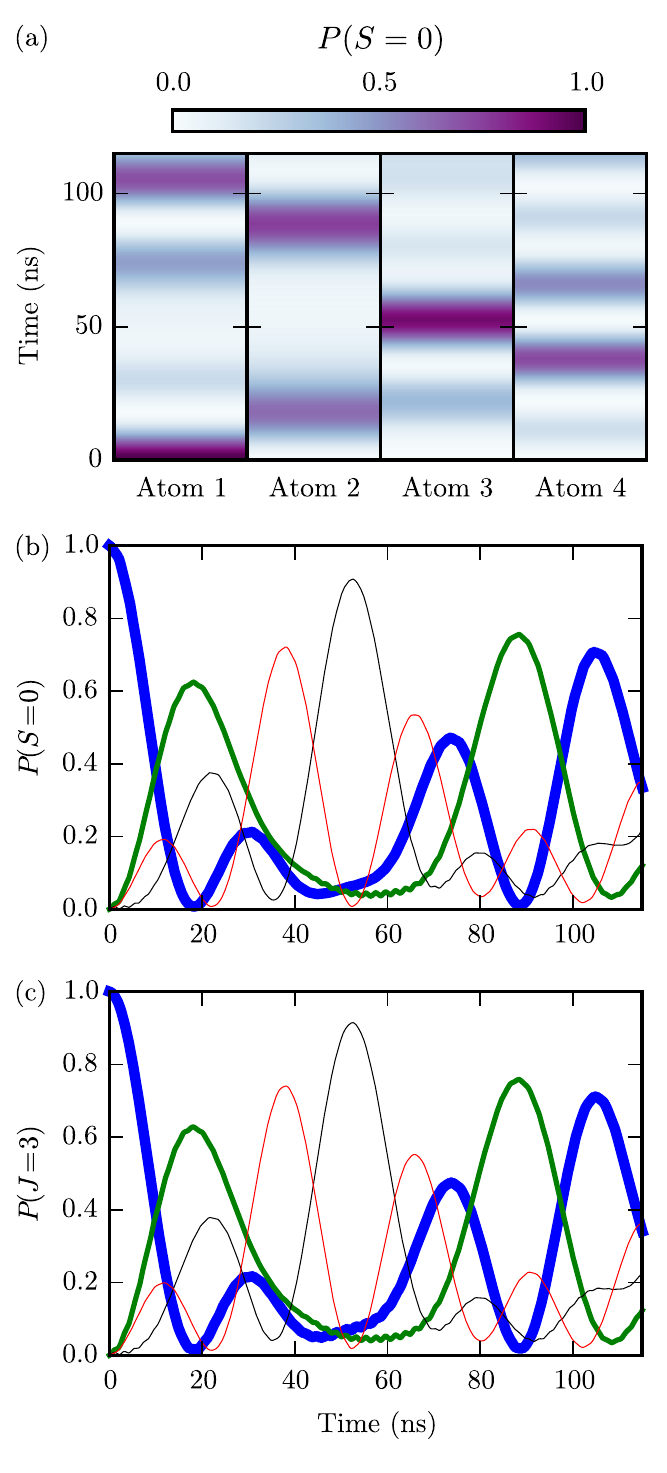}
\caption{\label{spinplotfixedR} (color online) (a) Spin-forbidden propagation of a singlet spin impurity ($28 \, ^1\!F_3$, initially located at Atom 1) along a short chain of triplet ($30 \, ^3\!D_2$) Rydberg atoms, with a spacing of $a=2.0 \; \mathrm{\mu m}$ between the atoms. Shading indicates the probability of each atom being in a $S=0$ state as a function of time. (b) and (c) Time evolution of the probability of finding the atom in a state with spin quantum number $S=0$ (b) and total angular momentum quantum number $J=3$ (c) for different atoms along the four-atom chain. Line styles denote atom 1 (thick blue), atom 2 (medium green), atom 3 (thin black) and atom 4 (thin red).}
\end{figure}

Since strontium has two valence electrons, two $5snd$ Rydberg series with total angular momentum $J=2$ exist, one labelled as the singlet and one as a triplet. Early theoretical and experimental work showed that these energy eigenstates do not have a well defined spin due to their interaction with doubly excited ``perturbers'' of mixed singlet/triplet character~\cite{Vaillant2014a, Esherick1977a, Wynne1977}. These perturbers are coupled to the Rydberg states by inter-electronic interactions, resulting in a breakdown of $LS$ coupling, which in turn affects the long-range inter-atomic interactions; the situation is depicted pictorially in Fig.~\ref{mainfigure}(a). We find that, for two interacting strontium atoms, the $|n \,^3\!D_2 ,\, n \,^3\!D_2 \rangle$ states are close in energy to the $|(n-2) \, ^1\!F_3,\, (n-3) \, ^3\!F_3\rangle$ states near $n=30$ (as shown in Fig.\ref{mainfigure}(b)), which, combined with the spin-mixing in the Rydberg series, allows near-resonant transfer between two-atom states. The impact of these intercombination near-degeneracies on the van der Waals coefficients is considered in Section \ref{dipolesection}. For the case of four atoms, we find that a spin impurity (a $28 \, ^1\!F_3$ atom) in a chain of $30 \, ^3\!D_2$ atoms can hop resonantly from site to site, showing a spin-forbidden propagation along the chain (as shown in Fig.~\ref{spinplotfixedR} and discussed in Section \ref{spinchainsection}).

\section{Multichannel Dipole-Dipole Interactions}
\label{dipolesection}
We begin by considering two atoms in $5snd \; ^3\!D_2$ states. Theoretically, this situation is most easily treated using multi-channel quantum defect theory (MQDT), which provides a wave function for each single-atom energy eigenstate in terms of a superposition of $LS$-coupled channels. Recently, we carried out an improved MQDT analysis of these states based on up-to-date experimental results, which gave the amount and nature of each electronic state (singlet, triplet perturber) present in each of the $J=2$ energy eigenstates~\cite{Vaillant2014a}. We use these wave functions to examine the long-range interaction between a pair of atoms prepared in the same $5snd \; ^3\!D_2$ energy eigenstate. Each of these pair states is coupled by electric dipole transitions to other final pair states (e.g. $|P, P\rangle$, $|P,F\rangle$, $|F,F\rangle$). Because of spin mixing, the final states may or may not differ in $S$ from the initial state. A key parameter is the energy difference between the final state and the initial state --- the so-called F\"orster defect --- which must be compared to the strength of the coupling.  Fig.~\ref{mainfigure}(b) shows an example where a near-degeneracy occurs in the spin-forbidden channel, i.e., where the spin labels of the initial and final states are different. Thus, although the spin mixing, and hence the coupling, is weak, this spin-forbidden process can become important.

More concretely, to describe the long-range interactions, we consider each atomic energy eigenstate state, $\Psi$, to be a sum over the MQDT channel states, $\phi_k$, such that $ \Psi = \sum_k \bar{A}_k \phi_k \chi_k$ (where $\chi_k$ is a function describing the angular, spin and remnant core state wave functions \cite{Vaillant2014a}). Using these state vectors, the long-range interactions can be calculated either perturbatively or non-perturbatively (by diagonalizing an effective Hamiltonian matrix in a basis of pair states~\cite{Vaillant2012}). The values of the coefficients $\bar{A}_k$, as well as numerical dipole matrix elements, are provided in~\cite{Vaillant2014a}. Throughout this paper, we only consider atoms that are initially in a stretched state ($J= |M_J|$), with the internuclear axes of the interacting atoms being aligned with the $z$-axis. Stretched states do not have any degeneracies in $M_{J_1} + M_{J_2}$, thereby reducing the number of states that need to be considered (even allowing for the fact that the dipole-dipole interaction couples stretched states to non-stretched states).

The dipole-dipole interaction Hamiltonian for two atoms with the internuclear axis aligned with the $z$-axis is given by~\cite{Dalgarno1966,Vaillant2012}
\begin{equation}
\begin{split}
H^{\mathrm{dd}} &= -\frac{4\pi}{3R^3} r_1 r_2 \left( Y_{1,1}(\hat{r}_1)Y_{1,-1} (\hat{r}_2)\right.\\
 &+ \left. Y_{1,-1}(\hat{r}_1)Y_{1,1} (\hat{r}_2) + 2 Y_{1,0}(\hat{r}_1)Y_{1,0} (\hat{r}_2) \right),
\end{split}
\label{dipoleinteraction}
\end{equation}
where $R$ is the interatomic distance, $r_1$ and $r_2$ are the radial electronic coordinates for atoms 1 and 2, $\hat{r}_1$ and $\hat{r}_2$ the angular electronic coordinates and $Y_{lm}$ denotes a spherical harmonic. The matrix elements of \eqref{dipoleinteraction} are thus products of angular factors and dipole matrix elements for each atom, with a $R^{-3}$ dependence.

Treating the dipole-dipole interaction using perturbation theory the first-order expression vanishes for two atoms in the same state. Second-order perturbation theory results in a $C_6 R^{-6}$ interaction, with~\cite{Vaillant2012}
\begin{equation}
C_6= R^6 \sum_i \frac{|\langle \Psi_1^{(i)} \Psi_2^{(i)}| H^\mathrm{dd} | \Psi^{(0)}_1 \Psi^{(0)}_2 \rangle|^2}{E^{(0)}_1 + E^{(0)}_2 - E_1^{(i)} - E_2^{(i)}},
\label{c6expression}
\end{equation}
where atom 1 is in state $| \Psi^{(0)}_1\rangle$ and atom 2 is in state $| \Psi^{(0)}_2 \rangle$ (note that these are not necessarily the same states), and the sum over $i$ runs over all the pair states dipole-coupled to $| \Psi^{(0)}_1 \Psi^{(0)}_2 \rangle$. By using the MQDT expansions of the wave functions in terms of channels and using the channel fractions and doubly excited state dipole matrix elements from~\cite{Vaillant2014a}, Eq. \eqref{c6expression} can be evaluated numerically.

\begin{table}
\centering
\caption{\label{c6table} The $C_6$ coefficients for the singlet and triplet $J=2$ $5snd$ ($|M_J| = 2$) configurations of Sr (where $n$ denotes the principal quantum number). The coefficients are given in atomic units, where $C_6 \mathrm{(GHz \; \mu m^6)}$ = $1.4448 \times 10^{-19} C_6$ (atomic units).}
\begin{ruledtabular}
\begin{tabular}{rdd}

$n$ & \multicolumn{1}{c}{$^1\!D_2$} & \multicolumn{1}{c}{$^3\!D_2$}\\
\hline
7&-7.98\times 10^7&-1.42\times 10^{8}\\
8&-1.29\times 10^{9}&6.79\times 10^{8}\\
9&-2.17\times 10^{10}&6.83\times 10^{8}\\
10&-2.95\times 10^{11}&1.03\times 10^{10}\\
11&-7.59\times 10^{10}&8.73\times 10^{10}\\
12&-9.53\times 10^{09}&4.03\times 10^{10}\\
13&-1.21\times 10^{13}&1.47\times 10^{12}\\
14&-3.37\times 10^{12}&3.54\times 10^{13}\\
15&-3.51\times 10^{12}&4.35\times 10^{12}\\
16&-3.83\times 10^{13}&-3.25\times 10^{12}\\
17&-2.51\times 10^{13}&7.58\times 10^{12}\\
18&-2.36\times 10^{13}&-4.47\times 10^{13}\\
19&-1.44\times 10^{13}&2.52\times 10^{14}\\
20&1.89\times 10^{13}&-3.48\times 10^{15}\\
21&1.05\times 10^{14}&-8.88\times 10^{14}\\
22&3.01\times 10^{14}&-6.83\times 10^{14}\\
23&7.01\times 10^{14}&-4.84\times 10^{14}\\
24&1.44\times 10^{15}&-2.44\times 10^{14}\\
25&2.77\times 10^{15}&1.92\times 10^{15}\\
26&5.12\times 10^{15}&4.26\times 10^{15}\\
27&9.04\times 10^{15}&9.86\times 10^{15}\\
28&1.55\times 10^{16}&2.34\times 10^{16}\\
29&2.61\times 10^{16}&7.37\times 10^{16}\\
30&4.21\times 10^{16}&4.08\times 10^{16}\\
31&6.66\times 10^{16}&1.41\times 10^{17}\\
32&1.03\times 10^{17}&2.92\times 10^{17}\\
33&1.57\times 10^{17}&5.83\times 10^{17}\\
34&2.35\times 10^{17}&1.15\times 10^{18}\\
35&3.46\times 10^{17}&2.43\times 10^{18}\\
36&5.03\times 10^{17}&5.59\times 10^{18}\\
37&7.22\times 10^{17}&1.70\times 10^{19}\\
38&1.02\times 10^{18}&-1.14\times 10^{21}\\
39&1.44\times 10^{18}&-3.51\times 10^{19}\\
40&1.99\times 10^{18}&-2.55\times 10^{19}\\
\end{tabular}
\end{ruledtabular}
\end{table}

\begin{figure}
\centering
\includegraphics{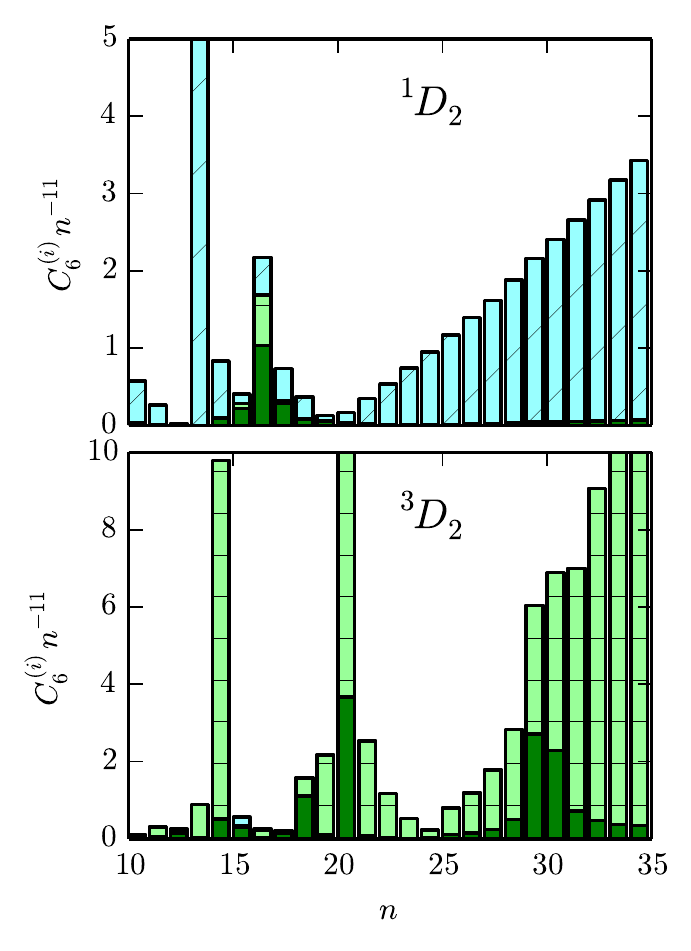}
\caption{\label{c6contributions} (color online) Absolute value of the contributions to the $C_6$ coefficients in atomic units (with the dominant $n^{11}$ scaling factored out for clarity) from the singlet-singlet (light blue diagonally-lined bars), singlet-triplet (dark green) and triplet-triplet (light green horizontally-lined bars) pair states acting as intermediate states. The initial states are taken to be in their stretched state, with $J=M_J$.}
\end{figure}

The resulting values of the $C_6$ coefficients for a pair of Sr atoms both in the same $5snd \; ^1D_2$ ($|M_J|=2$) state or the same $5snd \; ^3D_2$ ($|M_J|=2$) state are shown in Table~\ref{c6table}. The contributions from ``spin-allowed'' (i.e. singlet-singlet, triplet-triplet) and ``spin-forbidden'' (singlet-triplet) intermediate pair states are presented in Fig.~\ref{c6contributions}. Large contributions from singlet-triplet pair states are found in both series around $n=16$ where the effect of the $4d6s \; ^1\!D_2$ and $^3\!D_2$ perturbers is at its maximum~\cite{Vaillant2012,Esherick1977a}. The overall $C_6$ coefficients for states in this region differ significantly from predictions based on single-channel quantum defect calculations for Rydberg states below $n=30$~\cite{Vaillant2014b}. Above $n=30$, however, the calculated single-channel and multichannel values differ by less than $2\%$ of the overall $C_6$, thus validating the use of a one-electron treatment for high-lying Rydberg states of strontium~\cite{Millen2011,Vaillant2012,Zhi2001}.

Also visible in Fig.~\ref{c6contributions} is a large singlet-triplet contribution for $^3\!D_2$ states close to $n=30$. This arises due to the F\"orster resonance in the $ |n ^3\!D_2,n ^3\!D_2\rangle \rightarrow |(n-2) ^1\!F_3 , (n-3) ^3\!F_3\rangle$ channel shown in Fig.~\ref{mainfigure}(b). The uncertainties in the energy levels used to calculate the $C_6$ coefficients~\cite{Vaillant2012} are large enough that the location of the F\"orster resonance can change by one value of $n$; however the resonance is always present to within the error of these energy level measurements~\cite{Rubbmark1978, Esherick1977a, Vaillant2012}. The small F\"orster defect in this channel means that second-order perturbation theory breaks down at relatively large interatomic distances. We therefore turn to a non-perturbative calculation. Fig.~\ref{potentialcurves2atoms} shows the non-perturbative Born-Oppenheimer potential curves in the vicinity of the $|30 \,^3\!D_2 ,\, 30 \, ^3\!D_2\rangle$ asymptote, which has a spin-forbidden avoided crossing at relatively large distances ($R\sim 0.5 \; \mathrm{\mu m}$) with the $|28 \, ^1\! F_3 ,\, 27 \, ^3\! F_3\rangle$ asymptotic pair state. Without the mixing between the triplet and singlet series, the avoided crossing would not exist. While this F\"orster defect (522 MHz) is not small compared to that found in alkali atoms \cite{Walker2008, Singer2005, Reinhard2007}, the key point here is that it  is much smaller than the defect for the dipole-allowed pair states (8.84 GHz). As a result, the interaction between the spin-forbidden pair states is stronger than could be expected in view of the smallness of the singlet-triplet mixing in these Rydberg states.

\section{Spin Chain of Strontium Rydberg Atoms}
\label{spinchainsection}
\begin{figure}
\centering
\includegraphics{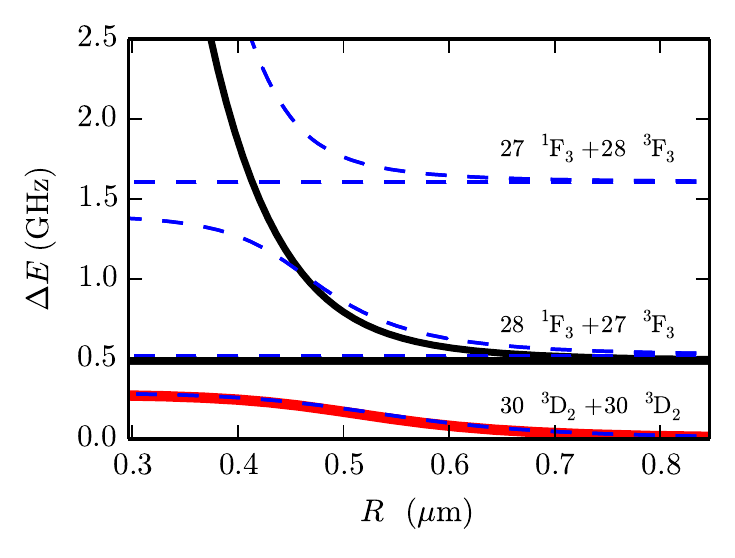}
\caption{\label{potentialcurves2atoms} (color online) Dipole-dipole potential curves for two atoms around the state labelled $|30 \, ^3\!D_2 ,\, 30 \, ^3\!D_2\rangle$ at infinity. The blue dashed curves represent the full non-perturbative calculation including all the relevant pair states, and the solid curves show the results of the four-level approximation (the red curve highlights the initial state, $|30\, ^3\!D_2,\, 30\, ^3\!D_2\rangle$). Only the $M_{J_1} = M_{J_2}=2$ states are shown for the $|30 \, ^3\!D_2 ,\, 30 \, ^3\!D_2\rangle$ state.}
\end{figure}

As another illustration of the impact of this intercombination F\"orster resonance on resonant energy transfer, we now examine the propagation of a singlet ``impurity'' in a short chain of four equally spaced atoms. While studies of state transport in lattices of Rydberg atoms have already been carried out~\cite{Wuster2010,Wuster2011}, as far as we know the impact of singlet-triplet mixing in this context has not been previously considered. Denoting the $30 \, ^3\!D_2$, $28 \, ^1\!F_3$, $27 \, ^3\!F_3$, and $28 \, ^3\!F_3$ states by $|0\rangle$, $|1\rangle$, $|2\rangle$, and $|3\rangle$, respectively, we numerically calculate the time evolution of the system at time $t$ after the $|1000\rangle$ state is prepared. Restricting the dynamics of each atom to these four states is justified by the fact that the $C_6$ coefficient of the $|30 \, ^3\!D_2,\, 30 \, ^3\!D_2\rangle$ is dominated by the F\"orster-resonant $|28 \, ^1\!F_3 ,\, 27 \, ^3\!F_3\rangle$ and the non-resonant $|28 \, ^3\!F_3 ,\, 27 \, ^3\!F_3\rangle$ pair states. All other pair states contribute less than $15\%$ to the $C_6$ coefficient of the $|30\, ^3\!D_2,\, 30\, ^3\!D_2\rangle$ state and are far enough away in energy to be neglected. Fig.~\ref{potentialcurves2atoms} shows the consequence of only choosing the four single atom states, $30 \, ^3\!D_2$, $28 \, ^3\!F_3$, $27 \, ^3\!F_3$, and $28 \, ^1\!F_3$. The potential curve for the $|30 \, ^3\!D_2 ,\, 30 \, ^3\!D_2 \rangle$ asymptote is well reproduced. We include all values of $M_J$ that contribute.

In order to perform the time-dependent calculation of the spin chain, we write the total Hamiltonian as $H= H_0 + H_\mathrm{int}$, where $H_0$ is the Hamiltonian of the four atoms with infinite lattice spacing. The interaction Hamiltonian $H_\mathrm{int}$ can be written as
\begin{equation}
H_{\mathrm{int}} = \sum_{i,j} V^{ji} |\Psi_{1}^{(i)} \Psi_{2}^{(i)}\Psi_{3}^{(i)} \Psi_{4}^{(i)}\rangle\langle \Psi_{1}^{(j)} \Psi_{2}^{(j)} \Psi_{3}^{(j)} \Psi_{4}^{(j)}|,
\label{Natomint}
\end{equation}
where $i$ and $j$ label the many-body states and the matrix element $V^{ji}$ is given by
\begin{equation}
V^{ji} = \sum_{p<q} \langle \Psi_1^{(i)} \Psi_2^{(i)} \Psi_3^{(i)} \Psi_4^{(i)} | H_{p q}^{\mathrm{dd}} | \Psi_1^{(j)} \Psi_2^{(j)} \Psi_3^{(j)} \Psi_4^{(j)} \rangle.
\label{interactionmatel}
\end{equation}
Here $H_{p q}^{\mathrm{dd}}$ is the dipole-dipole interaction Hamiltonian given in \eqref{dipoleinteraction} between lattice sites $p$ and $q$.

As the dipole-dipole interaction is a time-independent perturbation, we expand the eigenstates of $H$, $|\Phi_\alpha (a,t)\rangle$, in terms of the four-atom states of the lattice with infinite spacing, such that $|\Phi_\alpha (a,t)\rangle =  \exp(i\epsilon_\alpha (a) t) \sum_j U_{\alpha}^{(j)}(a)|\Psi_1^{(j)} \Psi_2^{(j)}\Psi_3^{(j)} \Psi_4^{(j)}\rangle$, where $\epsilon_\alpha(a)$ are the eigenenergies of $H$. The initial state vector $|\Psi_1^{(0)} \Psi_2^{(0)}\Psi_3^{(0)} \Psi_4^{(0)}\rangle$ is projected from the basis of the bare pair states into the eigenbasis. The $\exp(i\epsilon_\alpha t)$ factors are then easily determined, and the final state vectors are projected back into the original basis. To calculate the probabilities of the spin chain to be in a state $i$, we use the square magnitude of the coefficients

\begin{equation}
  c^{(i)}(a,t)= \sum_{\alpha} \left( U_{\alpha}^{(0)} \right)^* \exp(i \epsilon_\alpha t) \, U_{\alpha}^{(i)}.
\end{equation}

Fig.~\ref{spinplotfixedR} shows the evolution of the probability of each atom being in a singlet state for a lattice spacing of 2 $\mathrm{\mu m}$ (a spacing that can be engineered using two crossed 1550 nm laser beams \cite{Mukherjee2011, Nelson2007}). The spin can be seen to propagate along the chain of atoms and back, although there is additional state transfer due to competing second-order interactions. The calculation shown in Fig.~\ref{spinplotfixedR} includes the interactions between all the atoms, not just nearest-neighbour interactions. Nevertheless, a clear propagation of a spin singlet state through the chain can be seen, a phenomenon that can only occur due to spin-mixing.

\begin{figure}[htb]
\centering
\includegraphics{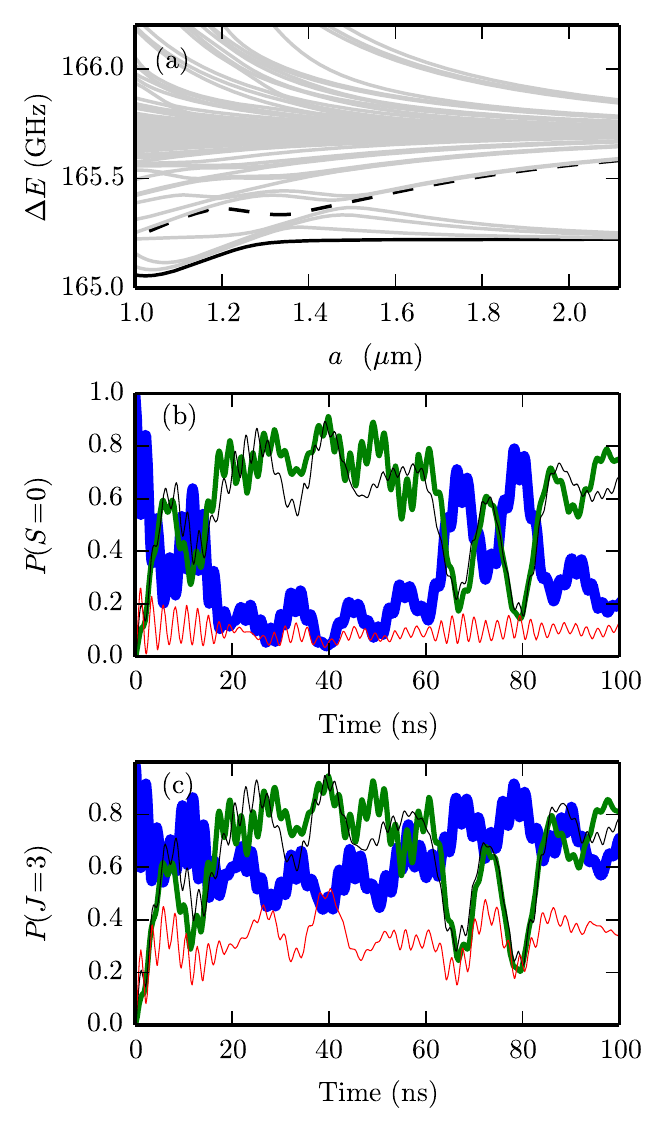}
\caption{\label{spinchainfigure} (color online) (a) Born-Oppenheimer energy curves for a spin chain of four atoms equally separated by a distance $a$ from their nearest neighbours, with the energy of one of the states of the $0,0,0,1$ family (solid black curve) and one of the states of the $0,1,1,2$ family (dashed black curve) highlighted for clarity. Subfigures (b) and (c) show the probability of atoms in the spin chain to be in a $S=0$ (b) and a $J=3$ state (c), for a lattice spacing of $a=1.35 \; \mathrm{\mu m}$. Line styles denote atom 1 (thick blue), atom 2 (medium green), atom 3 (thin black) and atom 4 (thin red).}
\end{figure}

For the parameters of Fig.~\ref{spinplotfixedR}, the dynamics arise primarily from the spin-forbidden dipole-dipole coupling between the four linearly independent states of the $0,0,0,1$ family, namely the states which reduce to linear combinations of the $|1000\rangle$, $|0100\rangle$, $|0010\rangle$ and $|0001\rangle$ states in the limit of infinite lattice spacing. As states 2 and 3 are unimportant here, there is essentially no difference between the probability for a particular atom of the chain being in a $S=0$ state, $P(S=0)$, and the probability of it being in a $J=3$ state, $P(J=3)$ (compare parts (b) and (c) of Fig.~\ref{spinplotfixedR}). However, this is not the case for smaller lattice spacings. As shown in Fig.~\ref{spinchainfigure}(a), the four $0,0,0,1$ states exhibit avoided crossings between 1.2 and 1.6 $\mathrm{\mu m}$ with the $0,1,1,2$ family of states, namely states in which, for $a \rightarrow \infty$, one of the four atoms is in state 0, two are in state 1 and one is in state 2. Due to these spin-forbidden F\"orster resonances and to the larger strength of the dipole-dipole interaction, the dynamics of the chain at $a=1.35 \; \mathrm{\mu m}$ is more complex than at $a=2 \; \mathrm{\mu m}$ [Figs.~\ref{spinchainfigure}(b) and (c)]. In particular, for some of the atoms the stronger coupling with state 2 at this smaller lattice spacing results in striking differences between $P(S=0)$ and $P(J=3)$. The key feature remains that the spin-orbit and inter-electronic effects responsible for the breakdown of $LS$ coupling \emph{within} each atom manifest spatially in the \emph{collective} state of the spin chain.

In regards to an experimental study of this intercombination dynamics, we note that the natural lifetimes of the $30\, ^3\!D_2$ and $28\, ^1\!F_3$ states (2.3 $\mathrm{\mu s}$ and 6.9 $\mathrm{\mu s}$, respectively~\cite{Vaillant2014a,Jonsson1984}) are much longer than its time scale. The triplet $F$ states lifetimes are unknown, but can be expected to be similar in magnitude to that of the $30\, ^3\!D_2$ and $28\, ^1\!F_3$ states. A possible experimental realisation could employ a strontium quantum gas microscope (QGM)~\cite{Weitenberg2011,Bakr2009}, where atoms are loaded into the lowest vibrational band of a 3-D lattice using a Mott-Insulator transition \cite{Stellmer2012}, and a high numerical aperture objective provides the required single-site readout. Single-site detection of Rydberg atoms in a QGM has already been demonstrated \cite{Schauss2012}. Since the atoms are in the lowest band, dephasing effects due to the uncertainty in the initial positions \cite{Ravets2014,Barredo2015,Ryabtsev2010,Ryabtsev2010b} are minimised.  Localisation to $<50$ nm is possible for reasonable lattice depths \cite{Weitenberg2011}, which is less than the width of the avoided crossing in Fig.~\ref{spinchainfigure}, and which is sufficient to observe coherent transport under the conditions of Fig.~\ref{spinplotfixedR}. We note that due to the fast timescales, the lattice could be switched off during the transport process. Finally, in order to image the state of the chain, short microwave pulses could be used to state-selectively transfer the population to other Rydberg states that do not interact resonantly, thus ``freezing'' the dynamics.

\section{Conclusions}
In conclusion, we have shown that intercombination transitions in Sr Rydberg atoms not only lead to a breakdown of $LS$ coupling but also allow dipole-forbidden excitation hopping along a chain of atoms via resonant long-range dipole-dipole interactions. We find that intercombination F\"orster resonances can have a substantial impact on long-range interaction. They can also lead to spatially separated dynamics between spin angular momentum and total angular momentum. Although we use Sr Rydberg states as an example, the ubiquity of spin mixing makes it likely that other systems may also show similar effects.

\acknowledgments{The authors would like to thank S A Gardiner and C W Weiss for useful discussions. Financial support was provided by EPSRC grant EP/J007021/1 and EU grant FP7-ICT-2013-612862-HAIRS. The data used in this publication can be freely downloaded from http://dx.doi.org/10.15128/cc24b379-1be2-4c42-862b-9760aa257077.}

\bibliography{spinchain}

\end{document}